# Electrical charging during the sharkskin instability of a metallocene melt.

S. Tonon, A. Lavernhe-Gerbier, F. Flores, A. Allal, C. Guerret-Piécourt*

Laboratoire de Physico-Chimie des Polymères, UMR CNRS 5067, 2 av du P. Angot, 64000 PAU

*Corresponding author: Christelle GUERRET LPMI-CURS BP1155 64013 PAU Cedex France, e-mail: christelle.guerret@univ-pau.fr, phone: 33. 5 59 40 77 08

**Abstract**

Flow instabilities are widely studied because of their economical and theoretical interest, however few results have been published about the polymer electrification during the extrusion. Nevertheless the generation of the electrical charges is characteristic of the interaction between the polymer melt and the die walls. In our study, the capillary extrusion of a metallocene polyethylene (mPE) through a tungsten carbide die is characterized through accurate electrical measurements thanks a Faraday pail. No significant charges are observed since the extrudate surface remains smooth. However, as soon as the sharkskin distortion appears, measurable charges are collected (around $5 \cdot 10^{-8}$ C/m$^2$). Higher level of charges are measured during the spurt or the gross-melt fracture (g.m.f) defects. This work is focused on the electrical charging during the sharkskin instability. The variation of the electrical charges versus the apparent wall shear stress is investigated for different die geometries. This curve exhibits a linear increase, followed by a sudden growth just before the onset of the spurt instability. This abrupt charging corresponds also to the end of the sharkskin instability. It is also well-known that wall slip appears just at the same time, with smaller velocity values than during spurt flow. Our results indicate that electrification could be a signature of the wall slip. We show also that the electrification curves can be shifted according to the time-temperature superposition principle, leading to the conclusion that molecular features of the polymer are also involved in this process.

**1. Introduction and bibliography**

On the one hand, flow instabilities occur during extrusion upon reaching a critical shear rate or stress. Due to their economical and theoretical interest, these phenomena have been extensively studied since their first report in the late 50's [1,2]. On the other hand, the electrification of polymers during their manufacturing process is also a well-known problem in industry, often resolved by the addition of antistatic agents. However few studies have been performed to correlate these both phenomena. Indeed among the hundreds of publications concerning the flow instabilities, only a tens of them are focused on the electrification of the polymer melt during their stable or instable flow [3-7].

The first study was performed by Vinogradov et al. on polybutadienes (PB) by using a cylindrical capacitor for the charge measurements under spurt and overspurt conditions. They have shown that a measurable electrification was only observed for pressure greater than the critical for spurt [3].

In the case of stable capillary extrusion of polyethylene (PE), polystyrene (PS) and polycarbonate (PC), Taylor et al. measured -by collecting the charges in a Faraday cup- an electrification attributed to the double layer formation, but no presence of flow instabilities were noted [4].

Ten years later the results of Dreval et al., - who has fulfilled the study begun by Vinogradov- on linear flexible chain polymers (PB, PI, fluorine rubber) did not confirm the electrification during the stable regime, indeed they only measured electrical charges during the spurt regime.

Very recently, the evidencing of electrification during the capillary extrusion of polymer has known a renewed interest, with the studies of Pérez-González et al. [7,8]. He used an experimental set-up where the charges are collected by swiping the probe of an electrometer on the molten extrudate. In these last papers no electrification during the sharkskin regime has been measured, since the appearance of the static electrification was linked to the conditions of strong slip.

All these previous studies have evidenced the influence of various parameters on the electrification of polymer melts such as the material of the die [4,7], the molecular characteristics of the polymer for instance the molecular weight [3], the molecular weight distribution, the segmental mobility [5], and the temperature [4]. Two mechanisms have been proposed for the electrification of the polymer during the extrusion, the first one is a tribological effect, the second one is a double layer effect.

In this paper, we focus on the electrification of a metallocene polyethylene (mPE) during the sharkskin defect occurrence. First, we will show that contrary to the studies of Vinogradov, Dreval and Pérez-González, electrical charges can be measured simultaneously with the presence of this defect. The fact that the wall slip appears at the same time as the electrical charging will be also underlined. Second, the effect of the die geometry on the electrification will be quantified, leading to some conclusions about the electrification mechanism. Third, the results at various temperatures will be presented and the time-temperature superposition principle will be applied to these curves. The whole study will show that rheological properties and electrical charging are strongly correlated.

## 2. Experimental section

*2.1. Material*

The studied polymer is a metallocene polyethylene (mPE) from Repsol. Its molecular characteristics are a weight average molecular weight of 118 000 g/mol and a polydispersity of 2.15. As shown by RMN study, the studied mPE is not branched, that is confirmed by the absence of strain hardening. Because of the polymerisation catalyst system, some metallic impurities are still present in the polymer melt. According to the safety data sheet of the mPE, their amounts were around 3.6 ppm for zirconium and 60 times more for aluminium.

The flow curve of this mPE is shown on figure 1 (a). It has been established in a capillary rheometer at a temperature of 190°C using a die of length over diameter L/D=20, D=2mm and a 180° entry angle. The sharkskin defect is the first occurring defect when the apparent shear rate $\dot{\gamma}_{app}$ reaches a value around 50 s$^{-1}$. It is replaced by the spurt flow defect at about 110 s$^{-1}$, and the last one is the gross-melt fracture (g.m.f) defect. In this paper, we will focus on the shear rate domain of the sharkskin defect.

To characterize the polymer, its elastic modulus G' and its viscous modulus G" have been studied on an imposed strain rheometer at different temperatures. The viscoelastic curves could be shifted horizontally and vertically according to the time-temperature superposition principle (see Fig. 1-b). The corresponding activation energy was 27.06 kJmol.$^{-1}$K$^{-1}$.

From the rheological master curve, the values of the Williams-Landel-Ferry (WLF) shift factors, that will be used later for charging phenomena, are $a_T$ = 1.753, and 1.363, for 155°C and 170°C respectively, with a reference temperature of 190°C and $b_T \approx 1$ in both cases.

One can note also on figure 1.b, that the cross-over appears for ω=80 rad/s, that corresponds to an average relaxation time for this polymer around 0.08 s at 150°C, or 0,05 s at 190°C.

*2.2. Experimental set-up*

The experimental set-up for measuring the electrokinetic charging of melt during the capillary extrusion is mainly composed of two coupled apparatus as shown on figure 2:

(i) The rheometer is a capillary extruder, where the polymer is first melt, then pushed by a piston at constant velocity through a die. The dies are made of tungsten carbide with an entry angle of 180° and the study is performed for dies of different length -L- over diameter -D- ratios -L/D-. Note that the minimum time spent by the polymer in the die, in the sharkskin zone, is about 1s at 150°C and 0,4s at 190°C.

(ii) The set-up for measuring the electrical charging is based on the measurement of the electric field induced by the charges present on the polymer extrudate. The mPE melt is collected, at the die exit, in a Faraday cup. The charges present on the extrudate induce the appearance of influence charge on the outside of the pail (equal and opposite to the nett quantity of charge placed into the cup). The electric field generated by these charges is then measured by a field-meter mill placed under the Faraday pail. The Faraday pail is a sensitive instrument for precise and reliable measurement of net electrostatic charge. This experimental set-up has a very good sensitivity that allows charges to be measured with a resolution down to 1pC -corresponding to surface charge down to $10^{-9}$ C/m$^2$-. The sensitivity is comparable to the one of the previous experimental set-up used by Taylor et al., that is based on the same principle of collecting the electrical charges [4].

## 3. Results on the electrification during extrusion

*3.1. Typical electrification curve*

One typical curve of electrical charging versus wall shear stress $\tau_w$ is presented on figure 3. Note that the charges remain zero until the extrudate is smooth. Then the density of charges increases regularly, with a charge density per unit area $\sigma$ in the range of $10^{-8}$-$10^{-7}$ C/m$^2$. The end of the sharkskin defect, that corresponds to the beginning of the spurt defect also coincides to a sudden increase of the electrical charging.

It is interesting to recall that the onset of wall slippage has been observed for such mPE before the onset of spurt [9,10]. This wall slip might be correlated to the appearance of the electrical charges.

But it has been impossible to obtain the velocity curve versus wall shear stress for the studied mPE by the classical Mooney method. Indeed the use of different die diameters has evidenced the variation of the flow curve with the die diameter (figure 3 (b)), that is generally considered as a proof of wall slippage. But, for the studied mPE, the measurements for three different diameters leads to some inconsistencies in the order of the flow curve versus the length diameter, so that the Mooney method is not reliable for determining the slip values, as previously noted by Robert et al. [11] . In this case, the slip velocity is probably not simply related to the wall shear stress, or the values of the slip are too small. The fact that wall slip is probably lower than the usual values observed for other mPE is consistent with the fact that this metallocene polymer, on the contrary to the one studied by Hatziriakos, is not branched [9,10]. Indeed Hatziriakos has explained the slip in terms of chains detachment from the interface, this detachment being favored by the branched feature of the mPE.

So the numerical link between $v_s$ and $\sigma$ could not have been established for this specific polymer but further work is in progress to evaluate the value of slippage for this mPE.

.

*3.2. Influence of the dies geometry*

The second point that has drawn our attention is the influence of the die geometry on the electrification. Figure 4 (a, b) shows the two curves of surface charge density σ versus wall shear stress $\tau_w$ for two die diameters D=1mm (Fig.4a), and D=2mm (Fig.4b). They sum up very well the observed trends : charging differs with both dies diameter and length. For example in the case of D=1mm, when increasing the length L of the die, the charge density σ seems to increase. On the contrary, for D=2 mm, increasing the length of the die seems to decrease σ, suggesting that the relationship between electrical charging and geometry of the die is not simple. But the fact that the charging rate is not linear versus dies length, at imposed diameter, is an important fact that will be discuss later.

A particular feature appears for the die with D=1mm and L=30mm where there is a sudden decrease that could be attributed to dielectric breakdown. Some dielectric breakdowns were also reported by Dreval et al. in spurt conditions, they attributed these breakdowns to the increase of the charge density with L, limited by the electric discharge processes between the polymer and the die wall at σ >4/6 $10^{-5}$ C/m$^2$ [6].

*3.3. Influence of the temperature*

The electrical charging measurements were performed for three temperatures of the mPE melt: T=155°C, T=170°C and T=190°C. On figure 5(a), for a die with L=30mm and D=2mm, one can observe that for these three temperatures, the variations of the charge density σ with the shear rate $\dot{\gamma}$ are the same and that the ranges of the charge density are also similar.

Note also, for experiment at T=170°C, that the initial charging occurs for $\dot{\gamma}_{i,170°C} = 20$ s$^{-1}$, and the final charging (corresponding to the beginning of spurt) for $\dot{\gamma}_{f,170°C} = 75$ s$^{-1}$, i.e. a range $\Delta\dot{\gamma} = 55$ s$^{-1}$. For T=155°C, the values are respectively $\dot{\gamma}_{i,155°C} = 32$ s$^{-1}$ and $\dot{\gamma}_{f,155°C} = 40$ s$^{-1}$, that corresponds to a range of $\Delta\dot{\gamma} = 8$ s$^{-1}$. For T=190°C, the values are $\dot{\gamma}_{i,190°C} = 60$ s$^{-1}$ and $\dot{\gamma}_{f,190°C} = 150$ s$^{-1}$ -i.e. a range of $\Delta\dot{\gamma} = 90$ s$^{-1}$-. So the electrical charging seems to follow the variations of the range of appearance of the sharkskin defect as shown on the cartography of the surface morphology Fig.5(c) established by Fernandez et al for the same mPE [12]. Indeed as the temperature decreases from 190°C to 170°C, the sharkskin defect arrives for smaller shear rate and the range of the sharkskin presence is reduced.

For electrical charging, in the case of T=155°C, the value of σ at low shear rate ($\dot{\gamma}_{i,155°C} = 32$ s$^{-1}$) is under-evaluate because, due to the low temperature of the melt, the extrudate became hard too quickly and could not be correctly stocked in the Faraday pail. But the same tendencies are still observed, the electrical charging occurrence corresponding to the sharkskin defect occurrence.

Analysis of the data from the figure 5(a) indicates that the thermo-dependence of the electrical charging is probably governed by an Arrhénius law like for all the linear viscoelastic functions. We assume that the activation energy for the electrical charging is the same as the one determined for the rheological G'-G" master curve building (cf. § 2.1), so that the same reduction parameters $a_T$ has been used to built the figure 5 (b). The resulting master curve for the electrical charging is correct, and this leads to an important conclusion: the time-temperature principle is applicable to the electrification process during polymer melt extrusion, that implies a contribution of the segmental mobility of the polymer in this electrification mechanism. By using the same horizontal shift factor, it has been also possible to built a master curve for L=20mm. An additional vertical factor would be necessary to obtain a good continuity of the master curve for L=10 mm.

**4. Discussion**

*4.1. Comparison with previous studies*

The comparison between these results and the previous studies evidences four main points :

First, for sharkskin instability, the charge density values are in the same range as the ones measured for laminar flow by Taylor. Indeed σ varies from $10^{-9}$ C/m$^2$ to $1.6 \cdot 10^{-7}$ C/m$^2$ as shown on figure 4, and Taylor measured charges between $10^{-7}$ C/m$^2$ to $10^{-5}$ C/m$^2$ for various polymers (PS, PC, LDPE...), the charges being the smallest for LDPE [4]. However this density of electrical charges is smaller than the one measured in spurt or overspurt conditions by Vinogradov et al., Dreval et al. and Pérez-González et al. Indeed they found charges around $10^{-5}$ C/m$^2$.

Second, it is the first time -to our knowledge- that electrical charges are observed during sharkskin defect. For Taylor study, they used a LDPE which did not exhibit sharkskin, and the other authors have noticed the presence of charge only for strong slip velocities under spurt and overspurt conditions. One explanation could be the better resolution of the experimental set up with a Faraday collector compared to the measurement by swiping the probe of the electrometer on the extrudate surface [7], or by the variation of a cylindrical capacitance [3]. If the correlation between electrical charges generation and slippage could be demonstrated, another explanation could be the fact that mPE generally exhibits higher slip values in the sharkskin domain as the other polymers [9].

Third this study has confirmed the influence of experimental parameters such as the length and the diameter of the die, but no clear variations of the charge density versus the surface area of the die could be evidenced.

Fourth, the influence of the temperature on the electrical charging has shown that the electrification is related to the molecular properties of the polymer such as the segmental mobility, since the time-temperature superposition is applicable.

*4.2. About possible mechanisms*

Two tentative explanations of the electrical charging can be proposed:

The first one is to imagine a mechanism like in solid-solid contact [13], named further triboelectrification mechanism. In the case of solids, the mechanism is well-known for the contact between two metals. Charges that are exchanged are electrons and the transferred quantity is proportional to the difference between the two work functions, that corresponds to the equality of the Fermi levels. Even in solid-solid contact, the charge transfer between a metal and an insulator is more complicated, because it depends on a lot of parameters: duration of the contact, number of contacts, area of contact... Even if no certain mechanism has been ascertained, most of the studies conclude that the exchanged charges are electrons and the quantity of transferred charges is governed by the difference between the metal Fermi level and some energy level characteristic of the insulator. In solid case, the transferred charges are usually in the range $10^{-5}$ to $10^{-3}$ $C/m^2$. It is also well-known that, for polymers, the determination of the real contact area is fundamental in the determination of the transferred charge density because of the increase of charge with time due to viscoelastic increase in the area of contact. For contact with sliding, it has been found that the charge deposited could be independent of the speed of sliding, or not, depending on the time of transfer and on the back-flow of charges [13]. The transfer of charge to a polymer has been supposed to be assisted by the stirring of molecules near to the surface, or by the number of polymer segments in contact with the metallic surface, and so by the molecular motion [14].

Similarly, in the case of a polymer melt going through a die, one can imagine that electrons are transferred between the polymer melt and the material of the die, and that the wall shear stress $\tau_w$ is mainly responsible of the number of contact between the polymer and the die wall, and so of the transferred charges. However in that case the quantity of transferred charge should be proportional to the length of the die at imposed diameter. But it has been shown previously on the experimental curves that the charging is not linearly linked to the die length. This contradiction induces two conclusions. Either the true contact area is not directly given by the surface of the die, which is likely for a viscoelastic material, or the hypothesis of the charging by electron transfer is not valid. Indeed, for an imposed die diameter D, one can suppose that the surface charge $\sigma_0$ exchanged between a surface unit of the die wall (located at the height x) and an infinitesimal disk of polymer (diameter D) is constant along the die ( from the entrance of the die (x=0) to the exit of the die (x=L),)

$$\sigma_0(x, D, T, \tau_{wx})=\sigma_0(0,D,T, \tau_{w0})$$

This assumption $\sigma_0(x, \tau_{wx})=\sigma_0(0, \tau_{w0})$ is based on the fact that the wall shear stress $\tau_w$ is mainly responsible of the transferred quantity of charges, and that in the present study, we can consider that $\tau_{wx} = \tau_w (x) = \tau_w (0) = \tau_{w0}$. Indeed, (i) the pressures used in this study have sufficient low values (P<500 bar) to consider that the influence of the viscosity is negligible on $\tau_w$, and (ii) the residence time of the melt in the die is much higher than the mean relaxation time of the mPE (see § 2.1 and 2.2), so that the flow can be considered to be established very quickly inside the die.

In that case of an exchange of charge constant along the die length, at imposed diameter and wall shear rate, the total surface charging of the infinitesimal disk of polymer might be directly proportional to the length of the die. So that the surface density of charges measured at the exit of the die might be linearly related to the die length.

If the charging by electron transfer is not valid, an alternative hypothesis might be the stripping of the well-known double layer. Indeed the tribocharging in liquids can be due to the relative motion between the fluid and the solid interface [4,15,16]. The charged metal surface attracts opposite charges and repels like charges. That leads to the formation of a double layer structure composed of an innerlayer formed of the adsorbed ions and of a diffuse layer (thickness $\lambda$) that compensates the innerlayer charge [17]. The properties of double-layers depend on the temperature, on the properties of the liquid and solid and also on the concentration of charged species in the liquid.

Under forced convection, two other layers can be defined, a stagnant layer (thickness $\delta_d$) near the interface and a flow layer further. For insulating liquids the thickness of the diffuse layer is greater than the thickness of the stagnant layer, $\lambda > \delta_d$, so that the electroneutrality is not recovered within the stagnant layer [16]. In that case, charges can be entrained in the flowing liquid. This theory, that has been initially developed for the electrostatic charging of oil during pipeflow, could be applied to the polymer extrusion [4,18]. Indeed charges are present in the polymer melt due for instance to metallic impurities in the case of the mPE (cf. § 2.1). The amount of ionic impurities necessary to built up an electrical double layer is extremely low (some ppm) [15]. This mechanisms fits well with the experimental facts observed for the tribocharging during sharkskin instability, i.e. the facts that the charging is linked to the molecular properties of the polymer, or to experimental parameters (length and diameter of the die, temperature...). Indeed parameters such as the diffusion coefficient of the liquid, the temperature, the flow velocity, die geometry etc, are directly linked to the tribocharging in this theory. Note also that, for newtonian fluids, $\delta_d$ is proportional to the reciprocal of the wall slip velocity, leading to an increase of the charging when the wall slip increases [15]. This point is consistent with our observation of a link between wall slippage and electrical charging. However in the case of mPE, further work is necessary to determine the nature of the ions present in the double layer, the influence of the slip velocity, and to adapt the published models to non-newtonian fluid...

Perhaps the both mechanisms are suitable successively. One can assume that the double layer stripping mechanism is dominant in the case of stable flow, and that the triboelectrification mechanism becomes dominant for unstable flow with strong wall slip.

## 5. Conclusion

It has been shown that the electrification of the melt is present simultaneously with the flow instabilities -in particular with the sharkskin-, its magnitude varying parallel to the nature of the instabilities.

The electrical charging behaviour is the same as the slip velocity behaviour according to the wall shear stress, that implies that the measurement of electrical charges could lead to an evaluation of the slip velocity -particularly for low slip velocity values-.

The study of the influence of experimental parameters leads to the conclusions that the electrification is not proportional to the apparent contact area between the melt and the die, and that the charging is related to the molecular properties of the polymer.

Finally two tentative explanations have been recalled for this electrical charge generation during extrusion: either it is due to an electron transfer between the polymer and the die like in solid-solid contact (named triboelectrification mechanism), or it is due to the stripping of the double layer due to the forced convection and the presence of ionic impurities in the polymer melt. Further work is necessary to choose between these two models or to determine the relative contribution of each model.

**Acknowledgments**

The authors acknowledge M. Fernández and A. Santamaría for the permission of using their results concerning temperature-shear map. This work was financially supported by the Conseil Régional d'Aquitaine (CTP).

**Figure captions**

Figure 1: (a) Flow curve of mPE performed in a capillary rheometer: T=190°C, length over diameter L/D=20, D=2mm and a 180° entry angle. (g.m.f = gross-melt fracture defect). (b) G' and G" curves for the studied mPE at T=150°C.

Figure 2: Schematic diagram of the apparatus

Figure 3: (a) Surface charge density $\sigma$ versus wall shear stress $\tau_w$, for mPE. (b) Flow curves for mPE obtained with capillaries of different diameters D at imposed L/D=10 at 190°C.

Figure 4: Surface charge density $\sigma$ versus wall shear stress $\tau_w$, at T=190°C, for various lengths of the die: (a) for a die diameter D=1mm, (b) for a die diameter D=2mm

Figure 5: Variation of the surface charge density $\sigma$ with the temperature, for L=30mm, D=2mm, (a) versus shear rate $\dot{\gamma}$, (b) versus $a_T\dot{\gamma}$, with the reduction parameters $a_T$ determine from the rheological study. (c): Temperature-shear rate maps for the surface morphology of the extrudate of the same mPE using a flat entry capillary tungsten die of 1mm diameter and a L/D ratio of 30. The borders are drawn on the basis of the experimental results marked by empty forms. This map has been established by Fernández et al., and is reproduced with permission of the authors,[12].

# References

[1] M. T. Dennison, Flow instability in polymer melts: a review, J. Plast. Inst. (1967) 803.

[2] M. M. Denn, Issue in viscoelastic fluid mechanics, Annu. Rev. Fluid Mech. 22 (1990) 13.

[3] G. V. Vinogradov, A. Y. Malkin, Y. G. Yanovskii, E. K. Borisenkova, B. V. Yarlikov G. V. Berezhnaya, Viscoelastic properties and flow of narrow distribution polybutadienes and polyisoprenes, J. Polym. Sc. 10 (1972) 1061.

[4] D. M. Taylor, T. J. Lewis T. P. Williams, The electrokinetic charging of polymers during capillary extrusion, J. Phys. D: Appl. Phys. 7 (1974) 1756.

[5] V. E. Dreval, V. P. Protasov G. V. Vinogradov, Linear flexible-chain polymer electrification conditions and pattern at T>TG, Polymer Bulletin 10 (1983) 343.

[6] V. E. Dreval, G. V. Vinogradov V. P. Protasov, The rheology and static electrication of linear flexible chain polymers, Proc. IX Intl. Congress on Rheology, Mexico,1984, Ed. Mena B (1984) 185.

[7] J. Pérez-González, Exploration of the slip phenomenon in the capillary flow of linear low-density polyethylene via electrical measurements, J. Rheol. 45 (2001) 845.

[8] J. Perez-González M. M. Denn, Flow enhancement in the continuous extrusion of linear low-density polyethylene, Ind. Eng. Chem. Res. 40 (2001) 4309.

[9] S. G. Hatzikiriakos, I. B. Kazatchkov D. Vlassopoulos, Interfacial phenomena in the capillary extrusion of metallocene polyethylenes, J. Rheol. 41 (1997) 1299.

[10] J. Pérez-González, L. de Vargas, V. Pavlínek, B. Hausnerová P. Sáha, Temperature-dependent instabilities in the capillary flow of a metallocene linear low-density polyethylene melt., J. Rheol. 44 (2000) 441.

[11] L. Robert, Instabilité oscillante de polyéthylènes linéaires: observations et interprétations, PhD Thesis, Université de Nice sophia-antipolis, France (2001) 146.

[12] A. Santamaria, M. Fernandez, E. Sanz, P. Lafuente A. Munoz-Escalona, Postponing sharkskin of metallocene polyethylenes at low temperatures: the effect of molecular parameters, Polymer 44 (2003) 2473.

[13] J. Lowell A. C. Rose-Innes, Contact electrification, Adv. Phys. 29 (1980) 947.

[14] K. Ohara, Contribution of molecular motion of polymers to frictional electrification, Inst. Phys. Conf. Ser. No. 48 (1979) 257.

[15] T. J. Harvey, R. J. K. Wood, G. Denuault H. E. G. Powrie, Effect of oil quality on electrostatic charge generation and transport, J. Electr. 55 (2002) 1.

[16] G. Touchard, Flow electrification of liquids, J. Electr. 51-52 (2001) 440.

[17] J. N. Israelachvili, Intermolecular and Surface Forces, Second edition, Academic Press, London, Great Britain, chap 12 (1992) 161.

[18] D. M. Taylor, Outlets effects in the measurements of streaming currents, J. Phys. D: Appl. Phys. 7 (1974) 394.

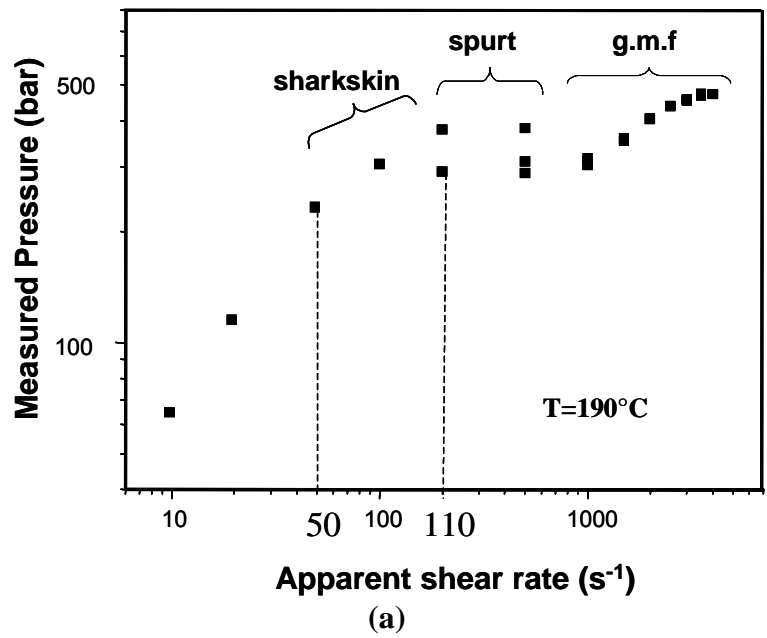

(a)

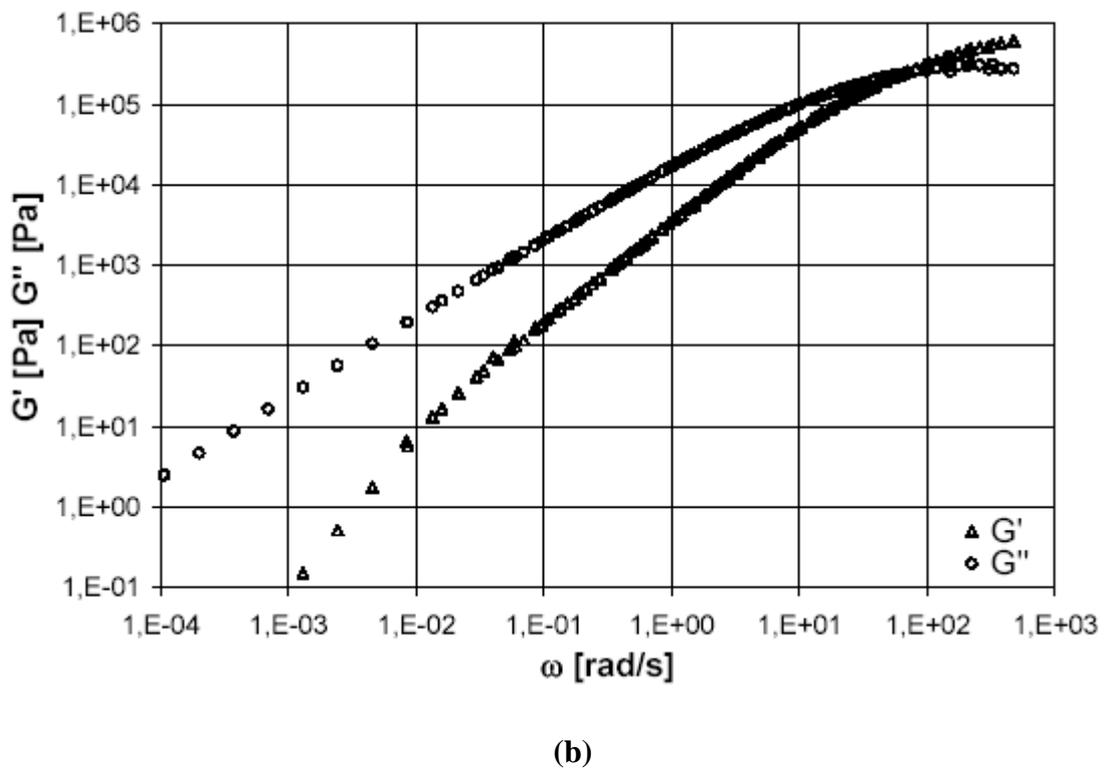

(b)

**Figure 1** S. Tonon et al., Electrical charging during the flow instabilities of a metallocene melt.

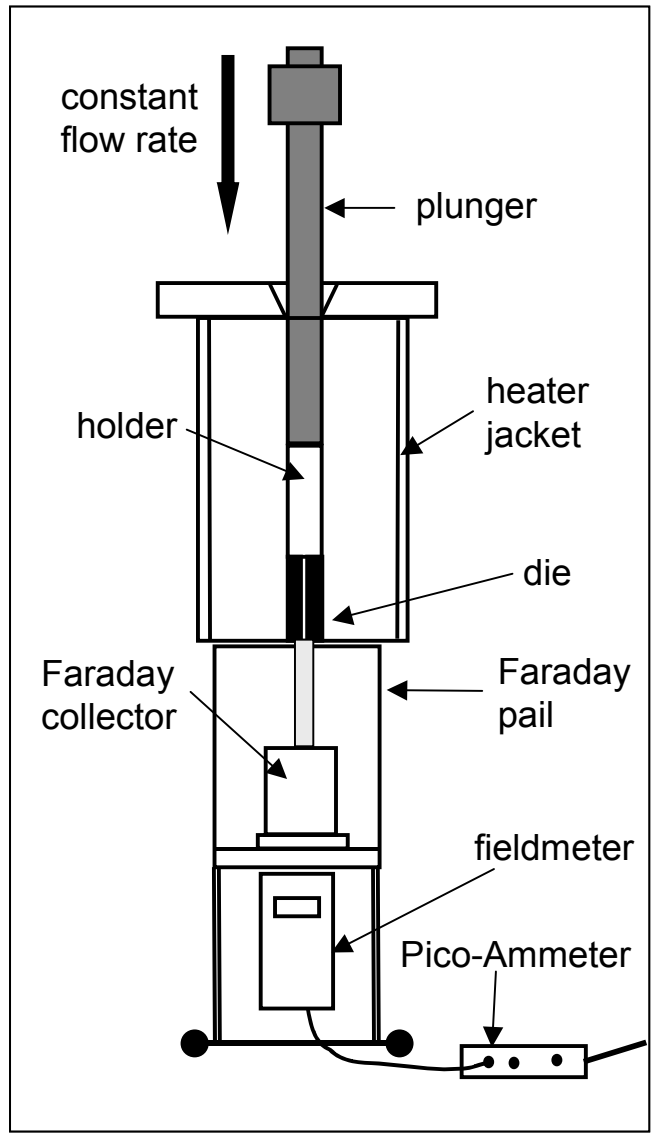

**Figure 2** S. Tonon et al., Electrical charging during the flow instabilities of a metallocene melt.

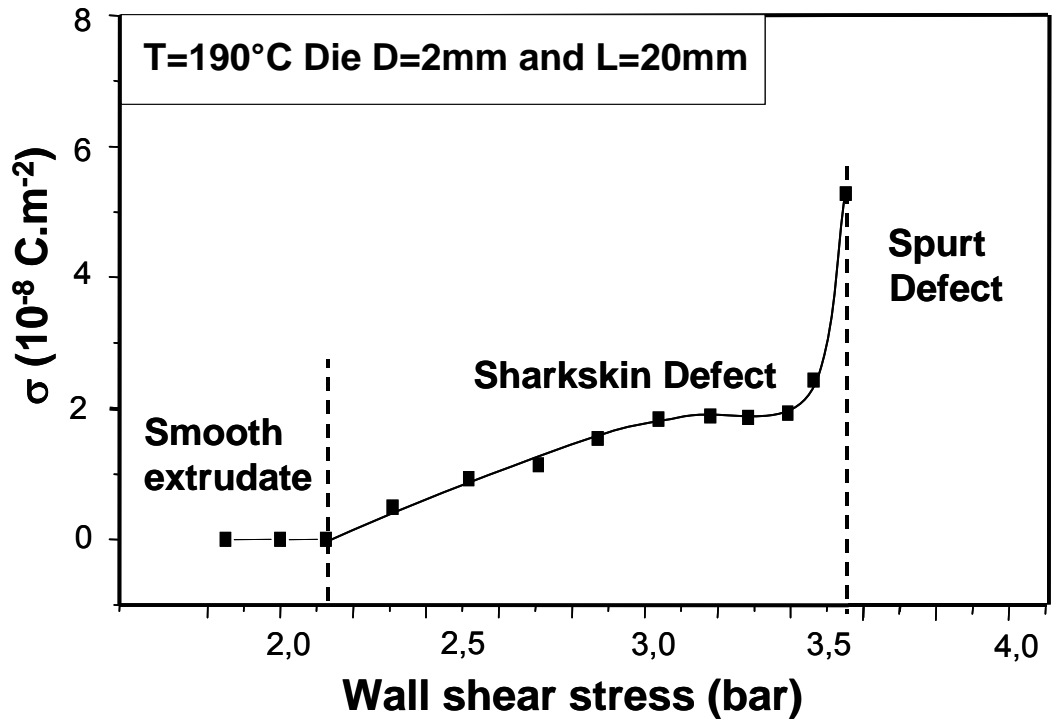

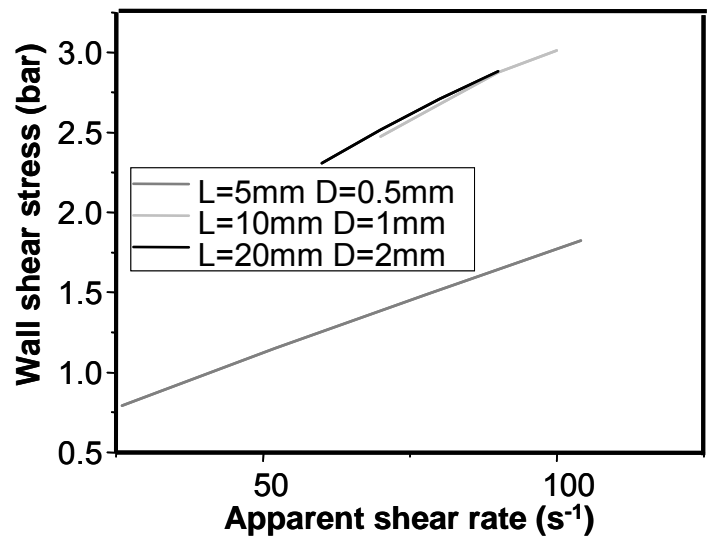

**Figure 3** S. Tonon et al., Electrical charging during the flow instabilities of a metallocene melt.

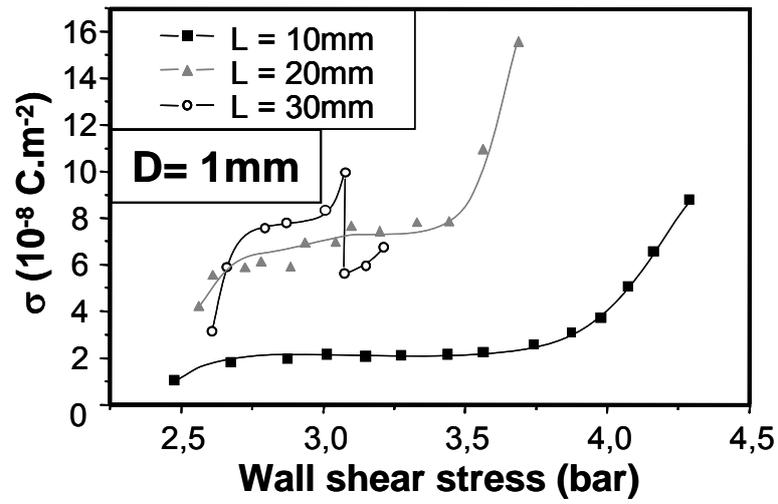

(a)

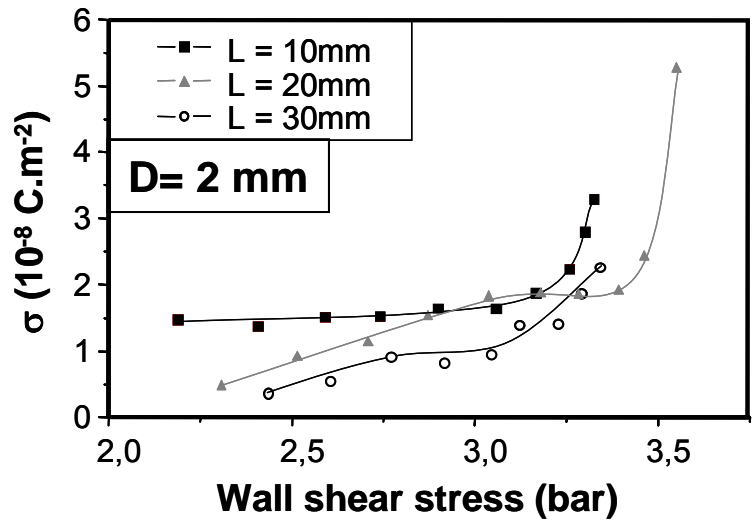

(b)

**Figure 4** S. Tonon et al., Electrical charging during the flow instabilities of a metal.locene melt.

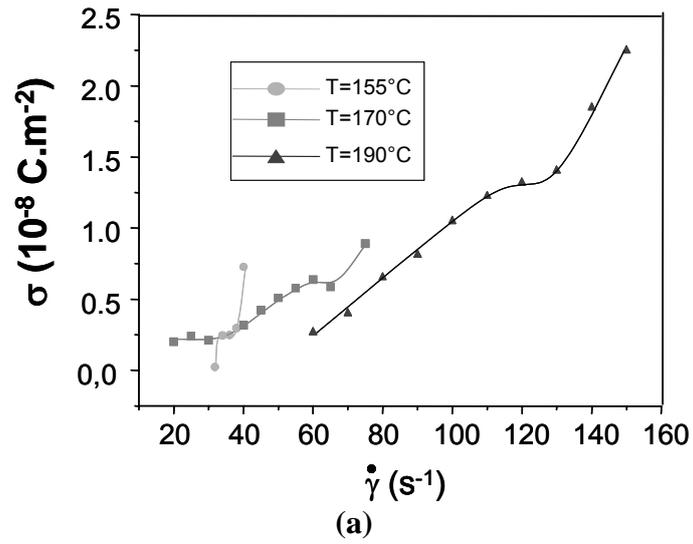

(a)

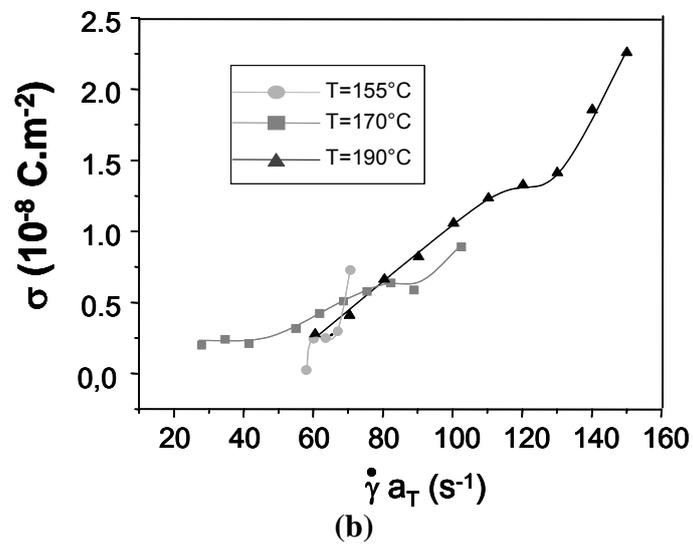

(b)

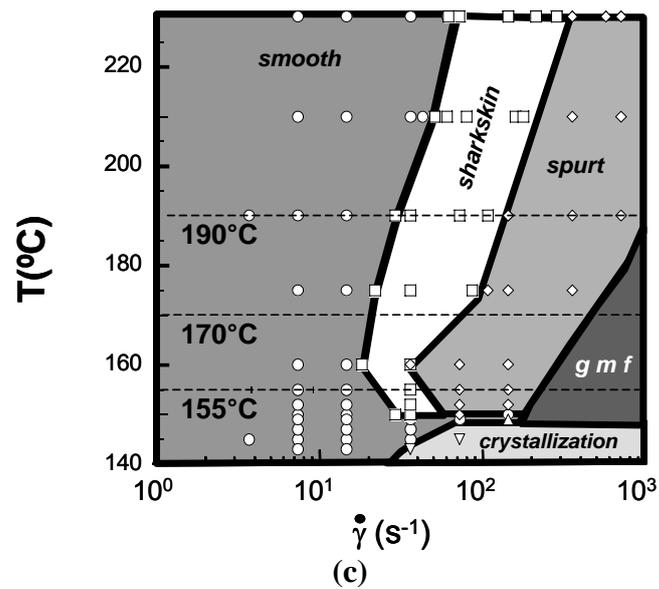

(c)

**Figure 5** S. Tonon et al., Electrical charging during the flow instabilities of a metallocene melt.

22